\documentclass[english,aps,prb,amsmath,floatfix]{revtex4}
\usepackage[T1]{fontenc}
\usepackage[latin1]{inputenc}

\usepackage{pslatex}
\usepackage{amsthm}
\usepackage{amsbsy}
\usepackage{pstricks}

\usepackage{babel}
\usepackage{natbib}
\usepackage{amsmath}
\usepackage{amssymb}
\usepackage{bm}
\usepackage[cspex,bbgreekl]{mathbbol}
\usepackage{bbm}
\usepackage{wasysym}
\usepackage{graphicx}  

\makeatletter

\providecommand{\LyX}{L\kern-.1667em\lower.25em\hbox{Y}\kern-.125emX\@}

\def\vec#1{\ifcat #1 a \mathbf{#1} \else \boldsymbol{#1} \fi }

\makeatother

\begin{document}
\newcommand{\hexa}{ \; \pspicture[0.35](1,0.5)
  \psdots[linecolor=black,dotsize=.15] (0.25,0)(0.25,0.5) (1,0.25)
  \psset{linewidth=0.03,linestyle=solid}
  \pspolygon[](0,0.25)(0.25,0)(0.75,0)(1,0.25) (0.75, 0.5) (0.25, 0.5)
  \psset{linewidth=0.03,linestyle=solid} \pspolygon[fillcolor=lightgray,
  fillstyle=solid](0.25,0.0)(0.75,0)(0.5,0.25) \pspolygon[fillcolor=lightgray,
  fillstyle=solid](0.25,0.5)(0.75,0.5)(0.5,0.25) \psline[](0.5,0)(0.5,0.5)
  \psset{linewidth=0.08,linestyle=solid} \endpspicture\;}

\newcommand{\hexb}{ \; \pspicture[0.35](1,0.5)
  \psdots[linecolor=black,dotsize=.15](0,0.25)(0.75,0) (0.75,0.5)
  \psset{linewidth=0.03,linestyle=solid}
  \pspolygon[](0,0.25)(0.25,0)(0.75,0)(1,0.25) (0.75, 0.5) (0.25, 0.5)
  \psset{linewidth=0.03,linestyle=solid} \pspolygon[fillcolor=lightgray,
        fillstyle=solid](0.25,0.0)(0.75,0)(0.5,0.25)
  \pspolygon[fillcolor=lightgray,
        fillstyle=solid](0.25,0.5)(0.75,0.5)(0.5,0.25) \psline[](0.5,0)(0.5,0.5)  

\psset{linewidth=0.08,linestyle=solid} \endpspicture\;}

\newcommand{\hexaf}{ \; \pspicture[0.35](1,0.5)
  \psset{linewidth=0.03,linestyle=solid}
  \pspolygon[](0,0.25)(0.25,0)(0.75,0)(1,0.25) (0.75, 0.5) (0.25, 0.5)
  \psset{linewidth=0.03,linestyle=solid} \pspolygon[fillcolor=lightgray,
  fillstyle=solid](0.25,0.0)(0.75,0)(0.5,0.25) \pspolygon[fillcolor=lightgray,
  fillstyle=solid](0.25,0.5)(0.75,0.5)(0.5,0.25) \psline[](0.5,0)(0.5,0.5)
  \psset{linewidth=0.08,linestyle=solid}
  \psdots[linecolor=black,dotsize=.15](0.25,0)(0.25,0.5) (1,0.25)(0.5,0.25)  
\endpspicture\;}

\newcommand{\hexbf}{ \; \pspicture[0.35](1,0.5)
  \psset{linewidth=0.03,linestyle=solid}
  \pspolygon[](0,0.25)(0.25,0)(0.75,0)(1,0.25) (0.75, 0.5) (0.25, 0.5)
  \psset{linewidth=0.03,linestyle=solid} \pspolygon[fillcolor=lightgray,
        fillstyle=solid](0.25,0.0)(0.75,0)(0.5,0.25)
  \pspolygon[fillcolor=lightgray,
        fillstyle=solid](0.25,0.5)(0.75,0.5)(0.5,0.25) \psline[](0.5,0)(0.5,0.5)
  \psset{linewidth=0.08,linestyle=solid}
  \psdots[linecolor=black,dotsize=.15](0,0.25)(0.75,0) (0.75,0.5)(0.5,0.25)
  \endpspicture\;}

\newcommand{\smallhexv}{\; \pspicture[0.35](0.2,0.4)
  \psset{linewidth=0.03,linestyle=solid}
  \pspolygon[](0.1,-0.1)(0,0.0)(0,0.2)(0.1,0.3)(0.2,0.20)(0.2,-0.0) 
\endpspicture \;}

\newcommand{\smallhexh}{\pspicture[0.35](0.4,0.2)
  \psset{linewidth=0.03,linestyle=solid}
  \pspolygon[](0,0.1)(0.1,0.0)(0.3,0)(0.4,0.1)(0.3,0.2)(0.1,0.2) 
 \endpspicture\;}

\title{On confined fractional charges: a simple model}

\author{F.~Pollmann and P.~Fulde}

\affiliation{Max-Planck-Institut f\"ur Physik komplexer Systeme,
  N\"othnitzer-Str.~38, 01187 Dresden, Germany}

\date{\today}

\begin{abstract}
We address the question whether features known from quantum chromodynamics
(QCD) can possibly also show up in solid-state physics. It is shown that
spinless fermions of charge $e$ on a checkerboard lattice with nearest-neighbor
repulsion provide for a simple model of confined fractional charges. After
defining a proper vacuum the system supports excitations with charges $\pm e/2$
attached to the ends of strings. There is a constant confining force acting between the fractional charges. It results from a
reduction of vacuum fluctuations and a polarization of the vacuum in the
vicinity of the connecting strings. 
\end{abstract}

\maketitle

One of the most fascinating aspects of quantum chromodynamics (QCD) is the
existence of particles with fractional charges, i.e., of quarks and their
confinement \cite{Gross73,Politzer73}. Of particular interest is that the
confining force is constant, i.e., independent of the distance of the separated
particles. The question may be posed whether this physical feature is
restricted to QCD or whether it occurs in a similar form in other areas of
physics, e.g., in solid-state theory. There are well known examples where parts
of the above mentioned physics show up in solids. We mention here that in
heavily doped polyacetylen excitations with fractional charges may exist
\cite{Su81}. In two dimensions the fractional quantum Hall effect is another 
example for the appearance of fractionally charged excitations
\cite{Laughlin83}. But in both cases there is no constant confining force. On
the other hand, confinement is nearly realized, e.g., in an Ising
antiferromagnet with a hole put into it. When this hole is moving with hopping
matrix element $t$ it generates a string of frustrated spins and therefore a
constant confining force - almost, one should add, because there exist
so-called Trugman paths \cite{Trugman88} which enable the hole to avoid
generating such string. The probability for those paths is small, though, i.e.,
the effective hopping matrix element is of order $t^5/J^4$ where $J$ is the
coupling constant of interacting neighboring Ising spins. But in this example
there are no fractional charges appearing.  

The aim of the present communication is to point out a simple Hamiltonian for
spinless (or fully spin polarized) fermions of charge $e$ defined on a
frustrated lattice, 
here a checkerboard lattice, which leads to excitations with fractional charges
$\pm e/2$ and a constant confining force. The advantage of studying such a
model system is that it enables us to describe in detail the origin of the
confining energy and constant confining force. We consider the following
Hamiltonian on a checkerboard lattice \cite{FuldeP02}

\begin{equation}
H = -t \sum_{\langle ij \rangle} \left( c^\dagger_i c_j + h.c. \right) + V
\sum_{\langle ij \rangle} n_i n_j
\label{eq:HtVij}
\end{equation}

\noindent where $c^\dagger_i(c_i)$ create (destroy) spinless fermions on site
$i$ and $n_i = c^\dagger_i c_i$. The notation $\langle ij \rangle$ refers to
neighboring sites. The
checkerboard lattice can be considered as a projection of a pyrochlore lattice
onto a plane. The latter consists of corner sharing tetrahedra and therefore
each of its sites has six nearest neighbors. We are interested here in the
limit $V \gg |t|$. 

{\begin{figure}
    \begin{center}
      \begin{tabular}{cc}
	(a)~\includegraphics[width=5cm]{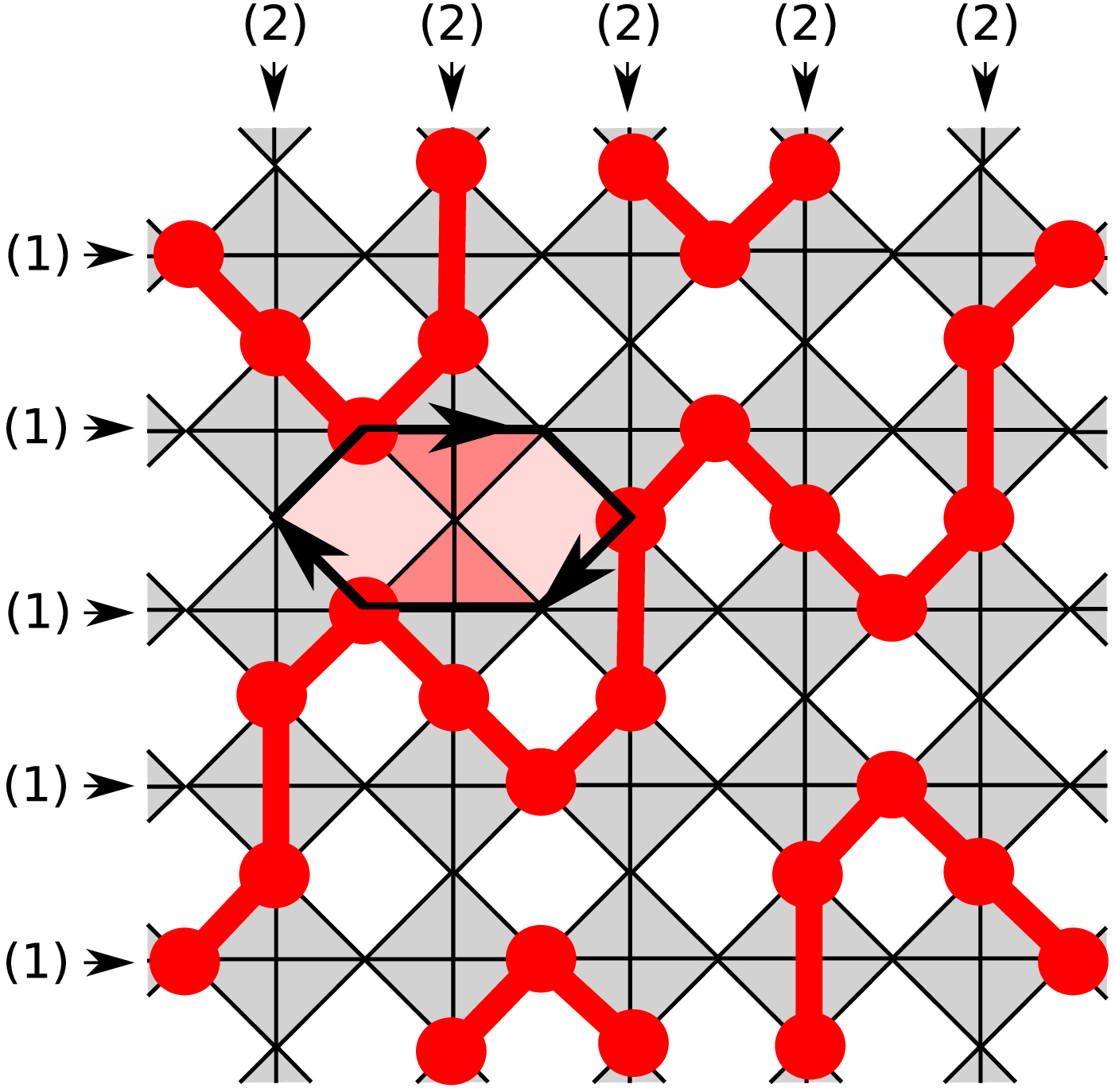}&
	(b)~\includegraphics[width=5cm]{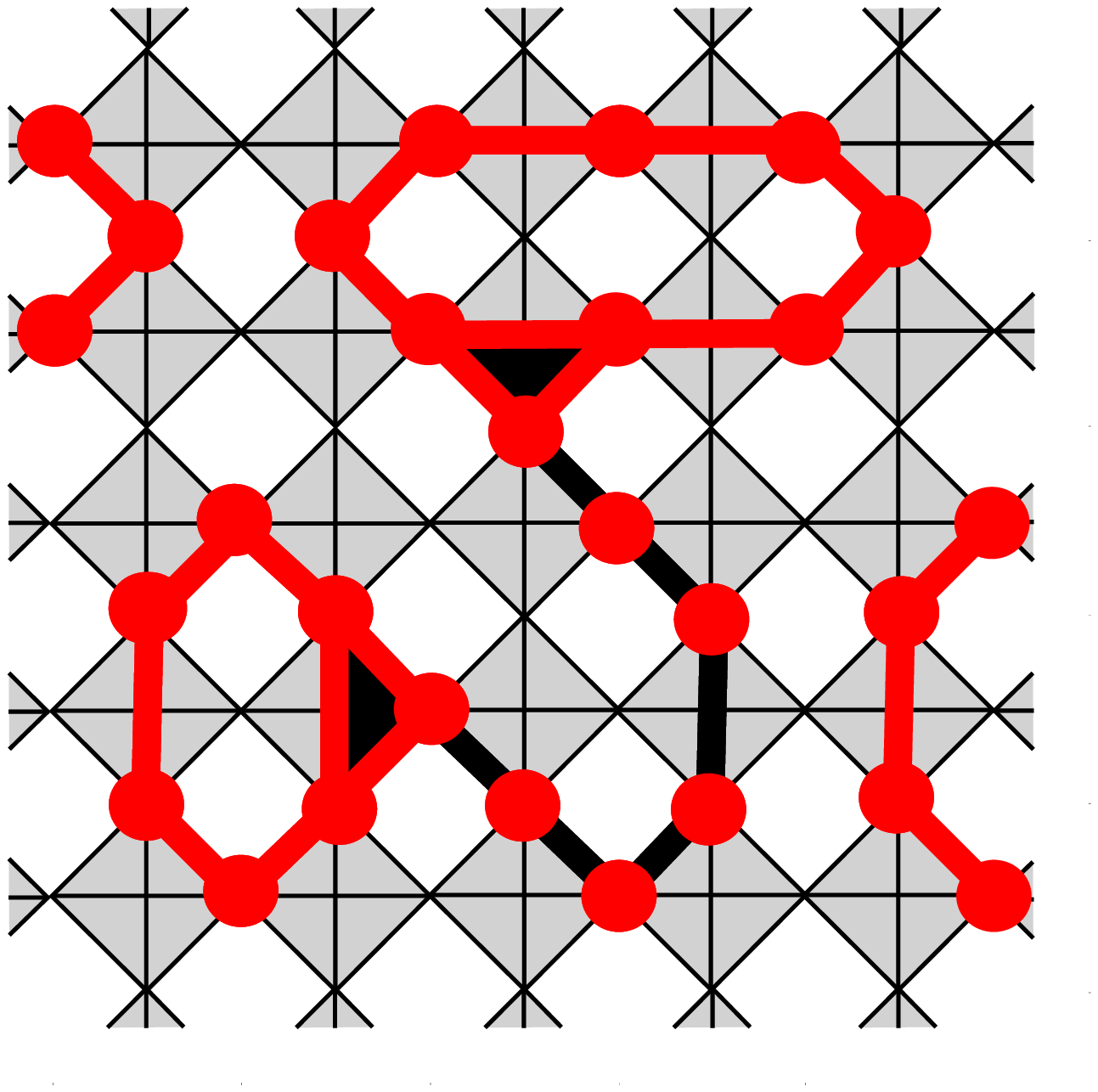}
	\tabularnewline
      \end{tabular}
    \end{center}
    \caption{(a) Ground-state configuration maximizing the number of flipable
      hexagons, one of which is explicitely shown. Neighboring occupied sites
      (dots) are connected by a solid line. (b) Two defects with charge $e/2$ of an
      arbitrarily chosen allowed configuration are connected by a string consisting
      of an odd number of particles, shown by the black line.
      \label{fig1}}
\end{figure}}

First we want to define a vacuum state for our system. For that
purpose we consider a lattice of $N$ sites with $N/2$ particles. When $t = 0$
the repulsion energy is minimized if every crisscrossed square is occupied
by two particles (''tetrahedron'' rule) \cite{Anderson56}. In that case the
(classical) ground state is approximately $(\frac{4}{3})^{\frac{3}{2} N}$ -
fold degenerate as known from the ice problem \cite{Pauling60}. One special of
those configurations $|\phi_i^{(0)}\rangle$ is shown in 
Fig. \ref{fig1}a. Here we have connected neighboring occupied sites by a solid
line. Assuming periodic boundary conditions those lines form loops which are
either contractible or go around the torus. The degeneracy of the ground states
is partially lifted by the hopping term \cite{Runge04}. To
lowest order in $t/V$ this occurs through ring hopping as described by the
effective Hamiltonian 

\begin{eqnarray}
H_{\mathrm{eff}} & = & \frac{12t^3}{V^2} \sum_{\smallhexh} c^+_{j_6} c^+_{j_4}
c^+_{j_2} c_{j_5} c_{j_3} c_{j_1}~~~. 
\label{eq:HeffFermions} 
\end{eqnarray}

\noindent It acts on the subspace spanned by the different ground-state
configurations and within that space on hexagons with sites $j_1, ... j_6$
which are alternating occupied and
empty. One of those hexagons is indicated in Fig. \ref{fig1}a. Note that ring
exchange is quite common in He$^3$ (see Ref. \cite{Thouless65}) and in
frustrated spin systems \cite{Diep05}. When matrix elements $\langle
\phi_i^{(0)} |H_{\mathrm{eff}}| \phi_j^{(0)}\rangle$ are calculated it matters
for the sign of the result whether the  site in the center of the hexagon is
occupied or empty. For that reason it is advantageous to rewrite
$H_{\mathrm{eff}}$ in the form 

\begin{eqnarray}
H_{\mbox{\tiny eff}} & = & -g\sum_{\{\smallhexh\}}\left(\Big | 
\hexaf\Big \rangle \Big \langle \hexbf\Big | -\Big | \hexa\Big \rangle \Big 
\langle \hexb\Big | +\mbox{H.c.}\right)
\label{eq:C4_H_eff}
\end{eqnarray}

\noindent with $g = 12 t^3/V^2$. The sum is over all horizontal and vertical hexagons of the system. The remaining notation is self-explanatory. Note that terms of order $t^2/V$ yield identical contributions for all ground states and therefore do not lift their degeneracy. Exact diagonalization of finite clusters with different geometries shows that within the quantum-mechanical ground state $| \psi_0 \rangle = \sum_i \alpha_i | \phi_i^{(0)} \rangle$ those $| \phi_i^{(0)} \rangle$ will have largest amplitudes $\alpha_i$ which contain the largest numbers of flipable hexagons. An investigation of the configurations $| \phi_i^{(0)} \rangle$ of finite clusters with up to 1000 sites shows that the ones with the largest number of flipable hexagons form squiggles. The latter have been first identified by Penc and Shannon \cite{Penc06}. Their unit cell has 40 sites. From translation and rotation of the unit cell a corresponding number of  different configurations $|\phi_{i}^{(0)}\rangle$ can be generated which maximize the number $N_{\mbox{fl}}$ of flipable hexagons. One of them is shown explicitely in Fig. \ref{fig1}a. These configurations correspond to two different charge orderings which are related by a rotation of $90^{\circ}$ around a symmetry axis. This is seen by counting the number of particles along the lines denoted by 1 and 2 in Fig. \ref{fig3}a. The charge ratio in the two directions is $\langle n_{x}\rangle:\langle n_{y}\rangle=2:3$ and for the rotated configuration $3:2$.

\begin{figure}
\begin{center}
\includegraphics[width=7cm]{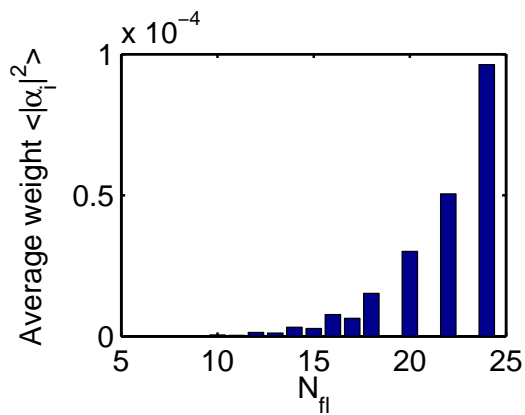} 
\caption{Average weights of configurations with $N_{\rm fl}$ flipable hexagons in the quantum mechanical ground state of a 72 site checkerboard cluster.
\label{fig2}}
\end{center}	
\end{figure}

When in a small cluster with periodic boundary conditions the squiggles can not develop, configurations with maximum $N_{\mbox{fl}}$
have a different shape. For quantum mechanical calculations we have used a cluster of size $\sqrt{72}\times\sqrt{72}$ with periodic boundary conditions. From the numerical analysis of such a cluster we have determined the distribution of the weight of configurations with given $N_{\mbox{fl}}$. The result is shown in Fig. \ref{fig2}. A single configurations $|\phi_{i}^{(0)}\rangle$ contributes to $|\psi_{0}\rangle$ the more, the more flipable hexagons it has. But due to their large number, the largest total contribution to the ground state of, e.g., a 72 site cluster comes from configurations $|\phi_{i}^{(0)}\rangle$ with 14 flipable hexagons, i.e., $N_{\mbox{fl}}=14.$ Since the squiggle phase cannot be realized in the considered cluster, we find a ratio $\langle n_{x}\rangle:\langle n_{y}\rangle=1:2$ and $2:1$ instead for the configurations which maximize $N_{\mbox{fl}}$. The Hamiltonian
(\ref{eq:C4_H_eff}) conserves that ratio. Since charge ordering is solely caused by the effective kinetic Hamiltonian (\ref{eq:C4_H_eff}), we have here a nice example of order by disorder. The quantum mechanical ground state is  therefore 2-fold degenerate. We define the vacuum by choosing one of the ground states, i.e., we are dealing with a symmetry broken (degenerate) vacuum. 

Next we place an additional particle onto an empty site of the checkerboard lattice. Hereby we select that particular ground state which in the presence of the added particle gives the lowest energy. As a consequence two neighboring tetrahedra have three particles attached to them. When the hopping term of Eq. (\ref{eq:HtVij}) is applied the two tetrahedra separate. That does not cost any additional repulsive energy. Each of the two tetrahedra has a charge $e/2$ attached to it, if the charge of the added particle was
$e$. Separating the two tetrahedra with three particles generates a string consisting of an odd number of sites which is linked to two loops with an even number of sites each. This is shown in Fig. \ref{fig1}b for an arbitrarily chosen configuration. The fractional charges sit at those linking points. Next we determine the changes of the kinetic energy density due to the presence of the two charges $e/2$. This is done by keeping the two tetrahedra with the fractional charges fixed at ${\bf 0}$ and ${\bf r}$ and determining the ground state $\bar{\psi}_0 ({\bf 0},{\bf r})$ and its energy for a cluster of 72 sites. The latter can be decomposed into local contributions by calculating the following expectation value of the energy at site $i$

\begin{eqnarray}
\epsilon_i& = & -\frac g 6\sum_{\{\smallhexh | i \in 
\smallhexh\}}\Big\langle \bar{\psi}_0(\mathbf 0,\mathbf r)\Big| \left(\Big | 
\hexaf\Big \rangle \Big \langle \hexbf\Big | -\Big | \hexa\Big \rangle \Big 
\langle \hexb\Big | +\mbox{H.c.}\right)\Big | \bar{\psi}_0(\mathbf 0,\mathbf 
r) \Big \rangle~~~.
\label{eq:epsig6}
\end{eqnarray}

\noindent The sum is over all hexagons containing site $i$. The result is shown in Fig. \ref{fig3}a,b. One notices a decrease in kinetic energy in the region between the two fractional charges, i.e., along the connecting string. The reason is that due to the topological changes caused by the generated string the number of flipable hexagons positioned between the fractional charges is  decreased. This energy decrease is proportional to the length of the generated string and implies a constant confining force. The origin of that force is clear. It is due to a reduction of vacuum fluctuations caused by the topological changes in the vacuum when the fractional charges are separated. Consider a configuration with two defects of charge $e/2$ separated along a diagonal (see Fig. \ref{fig3}b). Then we find that the maximal number of flipable hexagons is reduced by $2d$, where $d$ is the number of crisscrossed squares between the defects. We have also determined the density changes caused by the separation of the fractional charges (see
Fig. \ref{fig3}c,d). They are obtained from the ground state in the presence of the two defects $|\bar{\psi}_0({\bf 0},{\bf r})\rangle$ by simply calculating

\begin{equation}
\delta n_i = \left< \bar{\psi}_0 ({\bf 0},{\bf r}) \mid n_i \mid \bar{\psi}_0
({\bf 0},{\bf r}) \right> - \left< \psi_0 \mid n_i \mid \psi_0 \right>~~~. 
\label{eq:deltani} 
\end{equation}

{\begin{figure} 
\begin{center}\begin{tabular}{cc}
(a)\includegraphics[width=50mm]{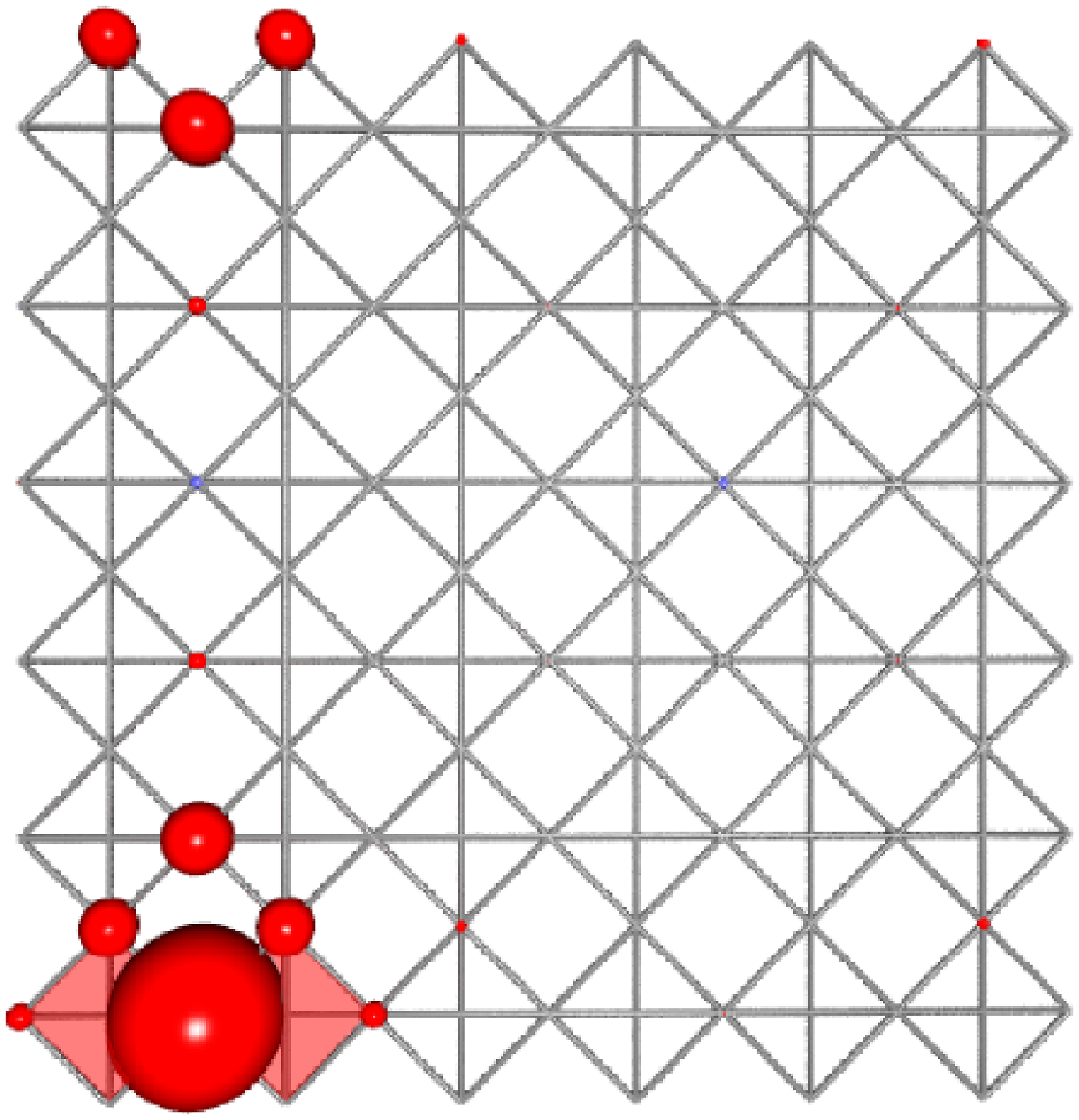}&
(b)\includegraphics[width=50mm]{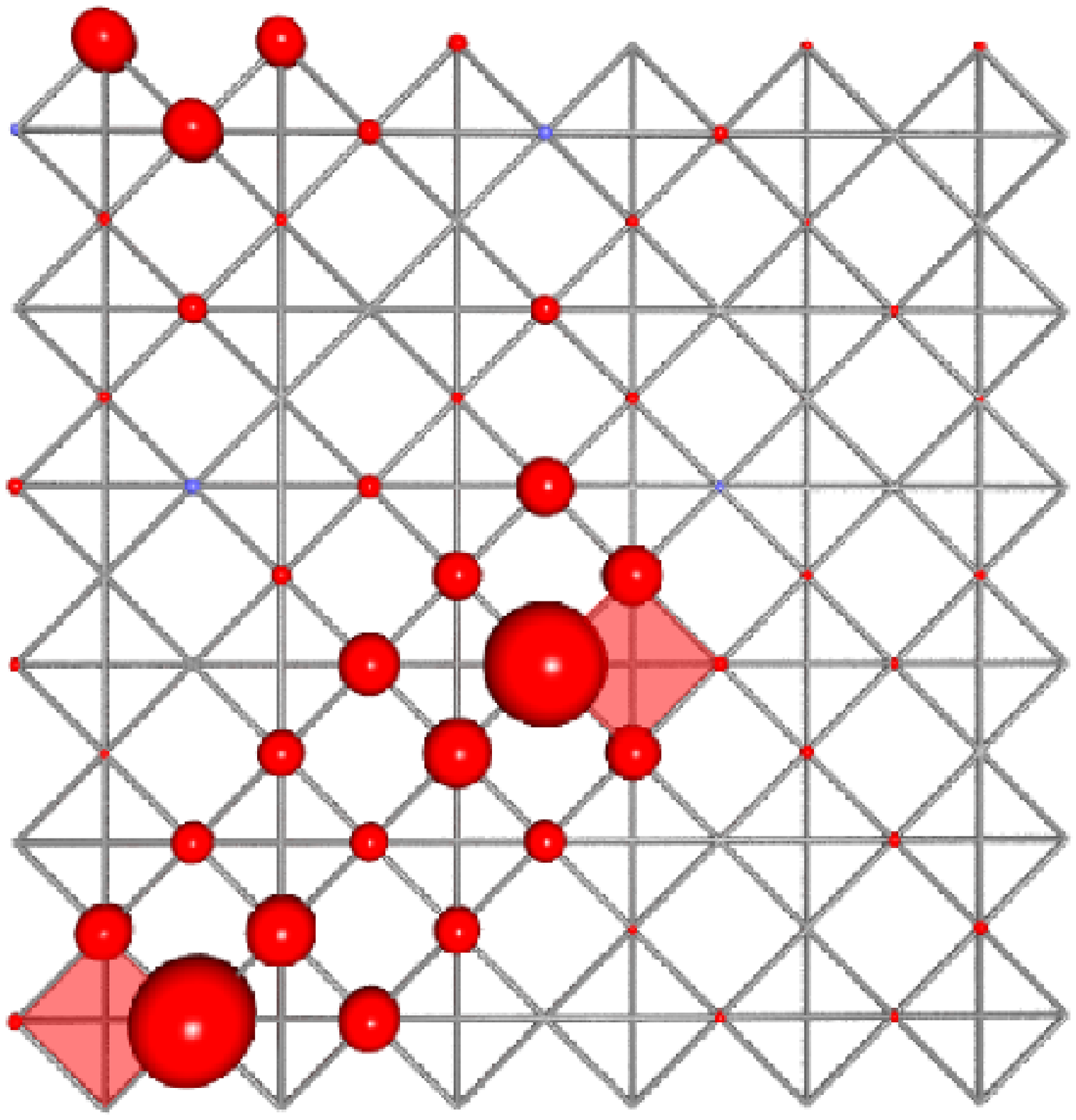}
\tabularnewline
(c)\includegraphics[width=50mm]{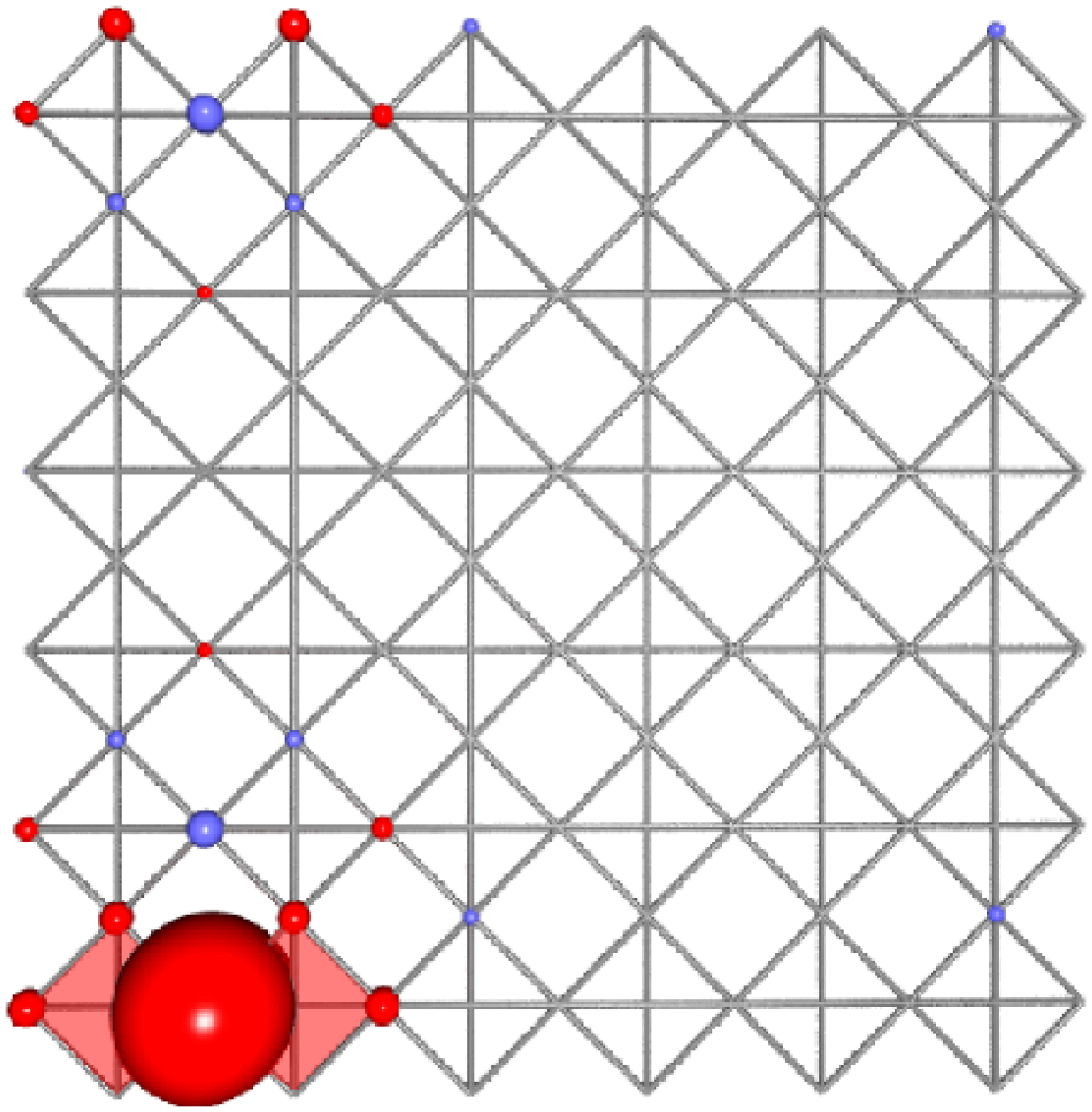}&
(d)\includegraphics[width=50mm]{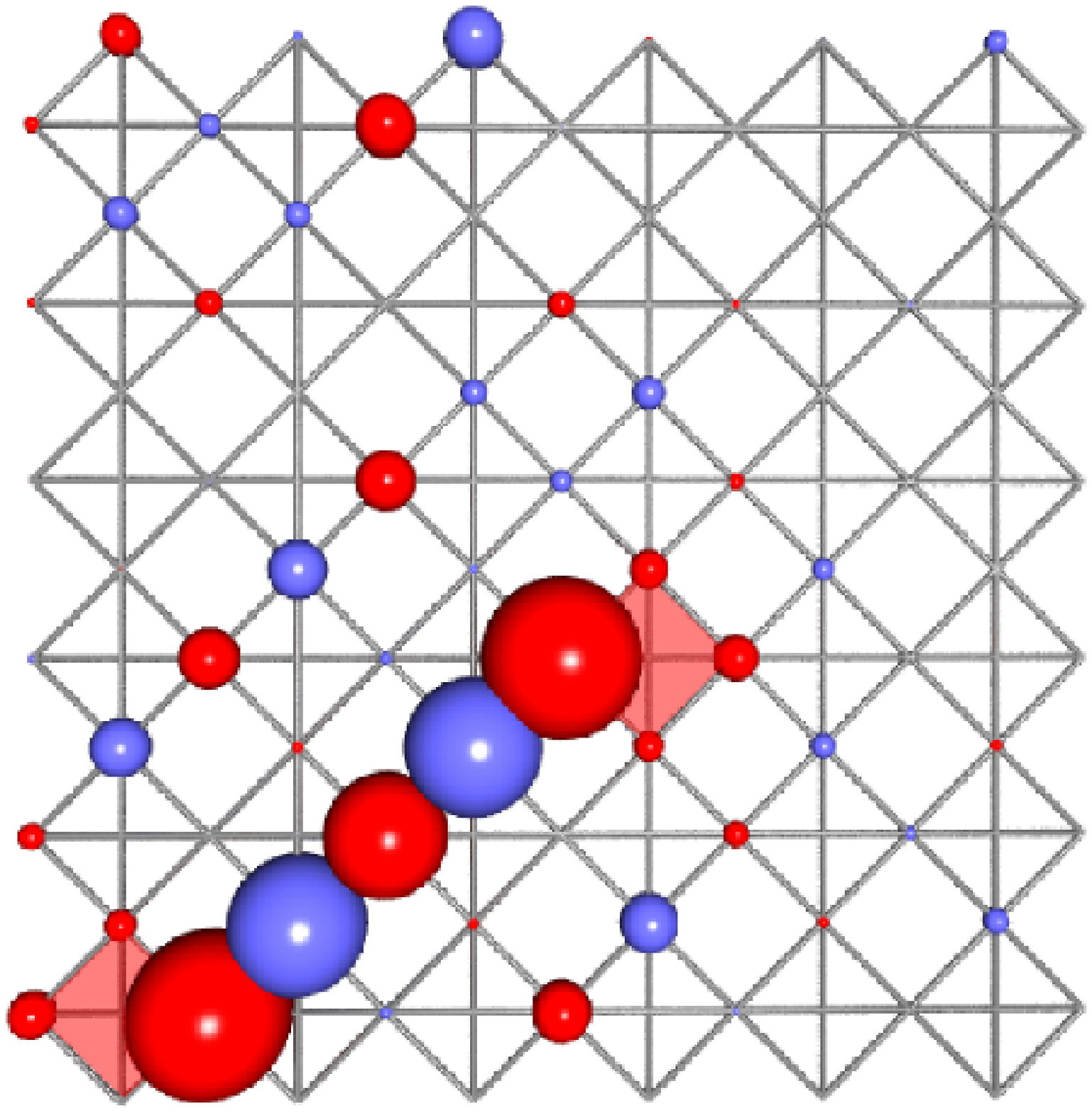}
\end{tabular}\end{center}
\caption{(a) and (b): local loss of kinetic energy due to a separation of two fractionally charged defects. Radii of the dots are proportional to the local energy loss. The positions of defects are shown by red colored crisscrossed square. (c) and (d): red and blue dots show an increase (decrease) of the local density (vacuum polarization).
\label{fig3}}
\end{figure}}

\noindent While the total changes add up to zero, the vacuum is modified by the breaking up of the charge. It is polarized along the connecting string. The above considerations can be also formulated in terms of a dimer model \cite{Rokhsar88}. For that purpose one connects the centers of the crisscrossed squares ending up with a square lattice. If the site on a link is occupied a dimer is attached to it. The tetrahedron rule implies that each site of the square lattice is touched by two dimers. 

The same findings discussed here apply also when a particle is removed from the ground state or when a particle-hole excitation is generated out of the ground state. They remain nearly unmodified for the case of quarter filling. We also expect to find a similar picture for the kagom\'e lattice, another frustrated lattice in which the classical ground-state manifold can be described by a dimer model on a bipartite lattice. We have not discussed here the three dimensional pyrochlore structure where also fractional charges may occur \cite{FuldeP02} because the numerics is much more involved. Neither have we discussed possible experimental verification of fractional charges in frustrated lattices \cite{Pollmann06}, e.g., in the rather common spinels (for the latter see, e.g.,
Ref. \cite{Tsuda00}). Instead we refer to the literature.  

Summarizing, we have found that a simple Hamiltonian (\ref{eq:HtVij}) when applied to a half-filled frustrated lattice leads in the strong coupling limit to fractional charges $\pm e/2$ with a constant confining force between them. The latter is due to modifications of the symmetry broken vacuum between the charges, i.e., the vacuum fluctuations are diminished and the vacuum is polarized along the string connecting the fractional charges. Our findings suggest that a number of features known from QCD are also expected  to occur in solid-state physics. Conversely, one would hope that by studying frustrated lattices or dimer models one might be able to obtain better insight into certain aspects of QCD. A crucial difference between the confinement in the introduced model and in QCD is the fact that we can adjust the confining force by varying the ratio $t/V$. For the considered limit $|t|\ll V$, the fractionally charged particles form bound pairs of large spatial extent \cite{Pollmann06}. Thus we expect that charge fractionalization has observable effects even if the fractionally charged particles are confined. For instance an exchange of fractionally charged particles might occur at finite doping.

\acknowledgments
We want to thank Dr. Karlo Penc (Budapest) and  Dr. Nic Shannon (Bristol) for sharing their own observations on the nature of the undoped ground state. Also discussions with Dr. Joseph Betouras (St. Andrews) on the mechanisms of confinement are gratefully acknowledged.

\end{document}